\documentclass[12pt]{article}
\usepackage{amsmath}
\usepackage[dvips]{graphicx}
\usepackage{epsfig}
\usepackage{amsmath}
\usepackage{amssymb}
\usepackage{graphics,epstopdf}
\usepackage{amsfonts}
\usepackage{amsthm}
\usepackage[usenames]{color}

\definecolor{AV}{rgb}{0.65,0.0,0}
\definecolor{GC}{rgb}{0,0.0,0.65}
\definecolor{WS}{rgb}{0,0.65,0}

\usepackage[
      colorlinks=true,
      linkcolor=blue,
      urlcolor=blue,
      filecolor=blue,
      citecolor=blue,
      pdfstartview=FitV,
      pdftitle={},
      pdfauthor={},
      pdfsubject={},
      pdfkeywords={},
      pdfpagemode=None,
      bookmarksopen=true
]{hyperref}

\newcommand{\bm}{\begin{multiline}}
\newcommand{\beq}{\begin{equation}}
\newcommand{\eeq}{\end{equation}}
\newcommand{\beqs}{\begin{eqnarray}}
\newcommand{\eeqs}{\end{eqnarray}}

\newcommand{\ra}{\rightarrow}

\begin{document}

\thispagestyle{empty}

\hfill{}

\hfill{}

\hfill{}

\vspace{32pt}

\begin{center}

\textbf{\Large  Charged black holes on the Taub-Bolt instanton }

\vspace{48pt}

\textbf{ Cristian Stelea,}\footnote{E-mail: \texttt{cristian.stelea@uaic.ro}}
\textbf{Ciprian Dariescu, }
\textbf{Marina-Aura Dariescu,}\footnote{E-mail: \texttt{marina@uaic.ro}}

\vspace*{0.2cm}

\textit{Faculty of Physics, ``Alexandru Ioan Cuza" University}\\[0pt]
\textit{11 Bd. Carol I, Iasi, 700506, Romania}\\[.5em]

\end{center}

\vspace{30pt}

\begin{abstract}

 We construct a new exact solution of the Einstein-Maxwell-Dilaton field equations in five dimensions, which describes a system of two general charged and static black holes sitting at the two turning points of the Taub-bolt instanton. We show that in this case the conical singularities can be completely eliminated and the black hole system remains in static equilibrium. We show how to recover some of the known solutions in particular cases and also obtain as a new solution the extremal double-black hole solution on the Taub-bolt instanton. Finally, we compute the conserved charges and investigate some of the thermodynamic properties of this system.  
\end{abstract}

\vspace{32pt}

\setcounter{footnote}{0}

\newpage

\section{Introduction}

There is no doubt that existence of black holes is one of the most important predictions of Einstein's General Theory of Relativity, since there is by now compelling evidence that such objects exist in Universe. In four dimensions, black holes were shown to obey the so-called `no-hair' theorem, which states that all regular asymptotically flat solutions of Einstein-Maxwell equations are uniquely determined by their conserved asymptotic charges, such as mass, angular momentum and electric charge; moreover they are given by the well-known Kerr-Newman class of solutions. That means that any dynamical black hole that eventually settles to a stationary state should actually be described by the Kerr-Newman solution.

 The higher dimensional Schwarzschild-Tangherlini black hole and its rotating generalization, the Myers-Perry solution have been long known \cite{Tangherlini:1963bw,Myers:1986un}, yet only recently, with Emparan and Reall's discovery of the asymptotically flat black ring solution in five dimensions \cite{Emparan:2001wn} (for a review of the physics of black rings see \cite{Emparan:2006mm}) one realized that the higher dimensional black holes exhibit a much richer behaviour than their four dimensional counterparts (for recent reviews on higher dimensional black holes see for instance \cite{Emparan:2008eg,Obers:2008pj}). The rotating black ring provided the first nontrivial example that known properties of the four-dimensional black holes do not hold in higher dimensions \cite{Emparan:2004wy}. Indeed, it was shown that in the case of four dimensional asymptotically flat space-times as a consequence of topological censorship \cite{Friedman:1993ty,Galloway:1999bp} all black holes should have spherical topology. However, in general dimensions, the spherical topology of infinity does not constrain that of the black hole horizon \cite{Galloway:1999br}. For instance, in five dimensions geometric considerations restrict the topology to those, such as $S^3$, $S^2\times S^1$ and lens-spaces, that admit non-negative scalar curvature \cite{Cai:2001su}. The black ring solution has the $S^2\times S^1$ topology of the event horizon, with rotation along the $S^1$ direction. Furthermore, the black ring can carry (in certain conditions) the same amount of mass and angular momenta as the spherical Myers-Perry black hole \cite{Emparan:2004wy}. Consequently, five-dimensional black holes are not uniquely characterized by their mass and angular momenta; the uniqueness theorems for black holes in four dimensions cannot be extended to the five dimensional case without further assumptions of additional symmetry and specification of the rod structure \cite{Hollands:2007aj}.

The derivation and the physical interpretation of the black ring solution was facilitated by the use of the so-called generalized Weyl formalism \cite{Emparan:2001wk,Harmark:2004rm}, which was specifically developed for geometries that in general $d$ dimensions admit $(d-2)$-commuting Killing vectors.  Following the discovery of the rotating black ring, its generalization to black Saturn \cite{Elvang:2007rd} and multi-black rings have been found in five dimensions \cite{Iguchi:2007is,Evslin:2007fv,Elvang:2007hs,Izumi:2007qx}. Concentric supersymmetric black rings in five dimensions were first constructed in \cite{Gauntlett:2004wh,Gauntlett:2004qy}. 

In five dimensions, there also exist the so-called squashed Kaluza-Klein (KK) black holes, whose horizon geometry is a squashed three-sphere \cite{Dobiasch:1981vh,Gibbons:1985ac,Rasheed:1995zv,Larsen:1999pp}. Their geometry is asymptotic to a non-trivial $S^1$ bundle over a four-dimensional asymptotically flat spacetime, which is also the asymptotic geometry of the Kaluza-Klein magnetic monopole \cite{Sorkin:1983ns,Gross:1983hb}. The black holes in such backgrounds look five-dimensional in the near-horizon region, while asymptotically they resemble four-dimensional objects with a compactified fifth dimension. Again, uniqueness theorems for KK black holes are proven assuming additional symmetry and specification of other invariants \cite{Hollands:2008fm}. KK black hole solutions  in presence of matter fields were generally found by solving the Einstein equations by brute force. For instance, a solution describing a static KK black hole with electric charge has been found in \cite{Ishihara:2005dp}, and the corresponding Einstein-Yang-Mills solution has been described in \cite{Brihaye:2006ws}. Remarkably, with hindsight, many  such KK solutions can be generated from known solutions by applying a `squashing' transformation on suitable geometries \cite{Wang:2006nw,Nakagawa:2008rm,Tomizawa:2008hw,Matsuno:2008fn,Tomizawa:2008rh,Stelea:2008tt}. However, not all of the KK black hole solutions can be generated by a squashing transformation; more general KK black holes have been derived  in the context of the minimal $5$-dimensional supergravity  \cite{Tomizawa:2008qr,Gal'tsov:2008sh,Mizoguchi:2011zj}. 

Even though the construction of vacuum rotating multi-black hole objects in five dimensions can be accomplished by use of the inverse-scattering method (see for instance \cite{Iguchi:2011qi} and references therein) one should note that in higher dimensions, by contrast to the single black hole case, solutions describing general charged multi-black hole systems are more difficult to derive. The main reason is that, except in the particular cases where the black holes are extremal/supersymmetric \cite{Myers:1986rx,Duff:1993ye,Ishihara:2006iv,Elvang:2005sa,Matsuno:2012hf,Matsuno:2012ge}, the known solution generating techniques lead to multi-black hole systems with charges proportional to their masses and, therefore, they cannot describe the most general charged solution for which the masses and charges are independent parameters. As is the case of the vacuum multi-black hole systems, progress in constructing charged multi-black hole solutions has been so far restricted to five dimensions. For example, an asymptotically flat solution describing a general double-Reissner-Nordstr\"om solution has been recently constructed in \cite{Chng:2008sr}, generalizing the uncharged solutions given in \cite{Tan:2003jz}. The solutions describing the general charged and static black bi-ring and di-ring systems in five dimensions have also been studied in \cite{Stelea:2011jm}. In spaces with KK asymptotics the general solutions describing charged multi-black hole solutions have been studied in \cite{Stelea:2009ur,Stelea:2011fj}.  For all these KK solutions, the background geometry, in absence of black holes is simply a direct product of a trivial time direction with the self-dual (multi-)Taub-NUT instanton. However, these are not the only possibilities for a background space-time geometry. Indeed, based on previous work \cite{Chen:2010zu}, in which they cast in Weyl form most of the gravitational instantons in four dimensions, Chen and Teo introduced in \cite{Chen:2010ih} a new class of five-dimensional black hole solutions, the so-called black holes on gravitational instantons. For these solutions, in absence of black holes, the background is again a direct product of a time direction with a four dimensional Ricci flat gravitational instanton. In particular, instead of the self-dual Taub-NUT instanton one could use the so-called Taub-bolt instanton, which is also an asymptotically locally flat gravitational instanton (similar to the Taub-NUT instanton) having a space like direction with finite norm at infinity. As an example, in \cite{Chen:2010ih} it was presented a new vacuum solution describing a black hole sitting on the Taub-bolt instanton. The thermodynamics of its charged version has been recently investigated in \cite{Nedkova:2011hx}.

 In this paper we are primarily interested in constructing a charged double black hole solution on background of the Taub-bolt instanton. The main motivation for this study was to see if one could find a static configuration describing two general charged black holes sitting at the turning points of the Taub-bolt instanton. Generically, in four and higher dimensions, it turns out that the solutions describing multi-horizon objects are plagued by unavoidable conical singularities. From a physical point of view, the presence of these conical singularities is easy to understand: they are needed to balance the gravitational attraction forces in between the black holes and also their electromagnetic interaction, in order to keep the system static. One way to balance the gravitational attraction is to add angular momenta to the system. This is what happens in the black ring case since rotation is needed there to keep the black ring from collapsing under its own gravity. In the same way one can balance the black Saturn system or the more general rotating bi/di-rings systems in five dimensions. Previous work \cite{Chng:2008sr,Stelea:2011jm,Stelea:2009ur,Stelea:2011fj} showed that, in absence of rotation, the electromagnetic interaction is not strong enough to balance the gravitational attraction in between the black holes. To our surprise however, it turns out that the double black hole on the Taub-bolt instanton can be equilibrated, even if the black holes are non-extremal. To our knowledge, this is the first example of a general charged static double-black hole system in equilibrium in five dimensions, without using Kaluza-Klein bubbles to balance the system.   
 
The structure of this paper is organized as follows. We first briefly recall the results of the solution generating technique given in \cite{Stelea:2009ur} that will allow us to lift four-dimensional charged static configurations to five dimensional Einstein-Maxwell solutions. In Section $3$, as a check of the method, we show how one can recover the charged single black hole solution sitting on the Taub-bolt instanton. As in \cite{Chng:2008sr,Stelea:2009ur}, in Section $4$ we shall use the general double Reissner-Nordstr\"om solution in four dimensions \cite{Manko:2007hi} as seed and lift it to five dimensions to obtain the desired double black hole system on the Taub-bolt instanton. The solution generating method extends easily to the more general case of Einstein-Maxwell-Dilaton (EMD) gravity with arbitrary coupling constant, however, for simplicity reasons, in this work we only consider the particular case of Einstein-Maxwell theory. In Section $4$ we also discuss some of the physical properties of this solution. Finally, we end with a summary of our work and consider avenues for future research.

\section{The solution generating technique}

Let us recall here the main results of the solution generating technique used in \cite{Stelea:2009ur}. The idea is to map a general static electrically charged axisymmetric solution of Einstein-Maxwell theory in four dimensions to a five-dimensional static electrically charged axisymmetric solution of the Einstein-Maxwell-Dilaton (EMD) theory with arbitrary coupling of the dilaton to the electromagnetic field. To this end one performs a dimensional reduction of both theories down to three dimensions and, after performing a mapping of the corresponding scalar fields and electromagnetic potentials of each theory, one is able to bypass the actual solving of the field equations by algebraically mapping solutions of one theory to the other. More precisely, consider a static electrically charged solution of the four-dimensional Einstein-Maxwell system with Lagrangian
\begin{eqnarray}  \label{4delectric}
\mathcal{L}_4&=&\sqrt{-g}\left[R-\frac{1}{4}\tilde{F}_{(2)}^2\right],
\end{eqnarray}
where $\tilde{F}_{(2)}=d\tilde{A}_{(1)}$ and the only non-zero component of $\tilde{A}_{(1)}$ is $\tilde{A}_{t}=\Psi$. The solution to the equations of motion derived from (\ref{4delectric}) is assumed to have the following static and axisymmetric form:
\begin{eqnarray}  \label{4dKhan}
ds_{4}^{2} &=&-\tilde{f}dt^{2}+\tilde{f}^{-1}\big[e^{2\tilde{\mu}}(d\rho
^{2}+dz^{2})+\rho ^{2}d\varphi ^{2}\big],~~~~~~~
{\tilde{A}_{(1)}} =\Psi dt.
\end{eqnarray}
Here and in what follows we assume that all the functions involved in our solutions depend only on coordinates $\rho$ and $z$. As shown in \cite{Stelea:2009ur}, the corresponding solution of the Einstein-Maxwell-Dilaton system in five dimensions with Lagrangian
\begin{eqnarray}
\mathcal{L}_{5}=\sqrt{-g}\left[R-\frac{1}{2}(\partial\phi)^2 -\frac{1}{4}e^{\alpha\phi}F_{(2)}^2\right]
\label{EMDaction5d}
\end{eqnarray}
where $F_{(2)}=dA_{(1)}$ can be written as:
\beqs
\label{final5dalpha}
ds_{5}^{2}&=&-\tilde{f}^{\frac{4}{3\alpha^2+4}}dt^{2}+\tilde{f}^{-\frac{2}{3\alpha^2+4}}\bigg[\frac{e^{2h}}{A^2-C^2e^{4h}}(d\chi+4ACH d\varphi)^{2}+(A^2-C^2e^{4h})e^{\frac{6\tilde{\mu}}{3\alpha^2+4}
+2\gamma-2h}(d\rho ^{2}+dz^{2})\nonumber\\
&&+\rho^2(A^2-C^2e^{4h})e^{-2h}d\varphi ^{2}\bigg],~~~~~~~
A_{(1)}=\sqrt{\frac{3}{3\alpha^2+4}}\Psi dt,~~~~~~~ e^{-\phi}=\tilde{f}^{\frac{3\alpha}{3\alpha^2+4}}.
\eeqs
Here $A$ and $C$ are constants, while $h$ is an arbitrary harmonic function\footnote{That is, it satisfies the equation $\nabla^2h=\frac{\partial^2h}{\partial\rho^2}+\frac{1}{\rho}\frac{\partial h}{\partial\rho}+\frac{\partial^2h}{\partial z^2}=0.$}, which can be chosen at will. Note that $h$'s presence can alter the rod structure of the final solution along the $\chi$ and $\varphi$ directions. By carefully choosing the form of $h$, one can construct the appropriate rod structures to describe the wanted configurations involving black holes sitting on the Taub-bolt instanton. Once the form of $h$ has been specified for a particular solution, the remaining function  $\gamma$ can be obtained by simple quadratures using the equations:
\begin{eqnarray}  \label{gammap1a}
\partial_\rho{\gamma}&=&\rho[(\partial_\rho h)^2-(\partial_z h)^2],~~~~~~~
\partial_z{\gamma}=2\rho(\partial_\rho h)(\partial_z h).
\end{eqnarray}
Also, the function $H$ is the so-called `dual' of $h$ and it is a solution of the following equation:
\beqs
dH&=&\rho(\partial_{\rho}h dz-\partial_zhd\rho).
\eeqs

Solutions of the pure Einstein-Maxwell theory in five dimensions are simply obtained from the above formulae by taking $\alpha=0$. For simplicity, in what follows, we shall focus on this case.

\section{Single black hole on the Taub-bolt instanton}

For a single black hole in five dimensions, the starting point in our solution generating technique will be the four-dimensional Reissner-N\"ordstr\"om solution, written here in Weyl form \cite{Emparan:2001bb}:
\begin{eqnarray}  \label{RN4dim}
ds^2&=&-\tilde{f}dt^2+\tilde{f}^{-1} \big[e^{2\tilde{\mu}}(d\rho^2+dz^2)+%
\rho^2 d\varphi^2\big], \\
\Psi&=&-\frac{4q}{r_2+r_3+2m},~~~~~~~\tilde{f}=\frac{(r_{2}+r_{3})^2-4%
\sigma^2}{(r_{2}+r_{3}+2m)^2},~~~~~~~ e^{2\tilde{\mu}}=\frac{Y_{23}}{2r_{2}r_{3}},  \notag
\end{eqnarray}
where we denote in general
\begin{eqnarray}
r_{i}&=&\sqrt{\rho^2+\zeta_i^2}, ~~~~~~~ \zeta_i=z-a_i, ~~~~~ Y_{ij}=r_ir_j+\zeta_i\zeta_j+\rho^2, 
\end{eqnarray}
while here $a_3=-\sigma$, and $a_2=\sigma$. Note that $\sigma=\sqrt{m^2-q^2}$, where $m$ denotes the mass and $q$ the electric charge of the black hole, while $\Psi$ is the electric potential in four dimensions.

To obtain the charged black hole sitting on the Taub-bolt instanton, one simply has to pick the following harmonic function:
\beqs
e^{2h}&=&\sqrt{\frac{r_2+\zeta_2}{r_3+\zeta_3}}\frac{r_1+\zeta_1}{r_2+\zeta_2}\equiv\frac{r_1+\zeta_1}{\sqrt{(r_2+\zeta_2)(r_3+\zeta_3)}}.
\eeqs
where $a_1>\sigma$. The first factor corresponds to a `correction' that has to be taken into account for each black hole horizon in the four-dimensional seed solution, while the second factor simply corresponds to a finite rod along the $\chi$-direction, which starts at $a_2$ and ends at $a_1$. By integrating (\ref{gammap1a}) one simply obtains:
\beqs
e^{2\gamma}&=&\frac{2^{\frac{3}{4}}}{K_0}\frac{\sqrt{Y_{12}{Y_{13}}}}{r_1}\frac{1}{\left(r_2r_3Y_{23}\right)^{\frac{1}{4}}},
\eeqs
where $K_0$ is a constant to be fixed later. The `dual' of $h$ can be easily found to be:\footnote{In general, the dual of $\frac{1}{2}\ln(r_i+\zeta_i)$ is $-\frac{1}{2}(r_i-\zeta_i)$, while the dual of $\frac{1}{2}\ln(r_i-\zeta_i)$ is $-\frac{1}{2}(r_i+\zeta_i)$, where $r_i=\sqrt{\rho^2+\zeta_i^2}$, $\zeta_i=z-a_i$ and $a_i$ is constant.}
\beqs
H&=&\frac{1}{4}\big[r_3-\zeta_3+r_2-\zeta_2-2(r_1-\zeta_1)\big]=\frac{1}{4}(r_2+r_3-2r_1)
\eeqs
up to a additive constant factor. Then the final solution in five dimensions can be written in the form:
\beqs
ds^2&=&-\tilde{f}dt^2+\frac{F}{\Sigma}(d\chi+\omega d\varphi)^2+\frac{\Sigma}{G}\big[e^{2\mu}(d\rho^2+dz^2)+\rho^2 d\varphi^2\big],\nonumber\\
A_{(1)t}&=&\frac{\sqrt{3}}{2}\Psi,
\eeqs
where we defined the following functions:
\beqs
F&=&\tilde{f}^{-\frac{1}{2}}e^{2h},~~~~G=\tilde{f}^{\frac{1}{2}}e^{2h},~~~~\Sigma=A^2-C^2e^{4h},~~~~e^{2\mu}=e^{\frac{3\tilde{\mu}}{2}+2\gamma},~~~~\omega=4ACH.
\label{functions}
\eeqs
So far the constants $A$ and $C$ were kept arbitrary. However, in order to have the right asymptotics at infinity, it turns out that one has to impose the following condition $A^2=C^2+1$. To show that we constructed the solution describing the charged black hole sitting on one of the turning points of the Taub-bolt instanton, let us consider for simplicity the uncharged version of this solution. If one sets $q=0$, then $\sigma=m$ and one obtains:
\beqs
\tilde{f}&=&\frac{r_2+\zeta_2}{r_3+\zeta_3},~~~F=\frac{r_1+\zeta_1}{r_2+\zeta_2},~~~G=\frac{r_1+\zeta_1}{r_3+\zeta_3}.
\eeqs

Let us consider the rod structure of this solution. Following the procedure outlined in \cite{Chen:2010zu}, note that there are three turning points that divide the $z$-axis into four rods:\footnote{We are writing the vectors in the basis $\{\partial/\partial t,  \partial/\partial \chi, \partial/\partial\varphi\}$.} 
\begin{itemize}
\item The first rod corresponds to $z<a_3$ and it has the normalized direction: 
\beqs
l_1&=&\sqrt{\frac{2\sqrt{2}}{K_0}}(0,AC(2a_1-a_2-a_3),1).
\eeqs
\item For $a_3<z<a_2$ one has a finite timelike rod that corresponds to the black hole horizon. Its normalized rod direction is given by $l_2=\frac{1}{k_E}(1,0,0)$, where 
\beqs
k_E&=&\left(\frac{K_0}{2\sqrt{2}A^2}\frac{1}{(a_2-a_3)(a_1-a_3)}\right)^{\frac{1}{2}}
\eeqs is the surface gravity on the black hole horizon represented by this rod.
\item For $a_2<z<a_1$ one has a finite spacelike rod with normalized direction $l_3=\frac{1}{k_{TB}}(0,1,0)$, where 
\beqs
k_{TB}&=&\frac{1}{2A^2}\sqrt{\frac{K_0}{2\sqrt{2}}\frac{1}{(a_1-a_3)(a_1-a_2)}}.
\eeqs
\item Finally, for $z>a_1$ one has an semi-infinite spacelike rod with normalized direction 
\beqs
l_4=\sqrt{\frac{2\sqrt{2}}{K_0}}(0,-AC(2a_1-a_2-a_3),1).
\eeqs
\end{itemize}
Take now $K_0=2\sqrt{2}$ and let us pick the values of $a_i$ such that $a_3=-c\kappa^2$, $a_2=c\kappa^2$, while $a_1=\kappa^2$, where $\kappa>0$ and $0\leq c<1$ are constants. Converting to the C-metric-like coordinates (see Appendix H of \cite{Harmark:2004rm}) one can easily see that one recovers the vacuum black hole solution on the Taub-bolt instanton found in \cite{Chen:2010ih}, provided one picks $A^2=\frac{1}{1-\alpha^2}$ and $C^2=\frac{\alpha^2}{1-\alpha^2}$. Note that $A^2=C^2+1$ as expected. Finally, to obtain a regular solution describing a static black one on the Taub-bolt instanton one has to take $\alpha=\pm\frac{\sqrt{1-c^2}}{2}$ such that the rod directions become $l_1=(0,2n,1)$, $l_2=\frac{1}{k_E}(1,0,0)$, $l_3=(0,4n,0)$ and $l_4=(0,-2n,1)$, where we defined the nut charge $n=\pm\frac{2\kappa^2\sqrt{1-c^2}}{3+c^2}$. They correspond to a black hole with surface gravity $k_E$ sitting on one of the two turning points of the Taub-bolt instanton.

In the charged case, taking $a_1=R$ and $K_0=2\sqrt{2}$ one obtains the normalized rod directions as follows: 
\beqs
l_1&=&(0,2n,1),~~~l_2=\frac{1}{k_E}(1,0,0),~~~l_3=\frac{1}{k_{TB}}(0,1,0),~~~l_4=(0,-2n,1),
\eeqs
where we defined $n=ACR$. Here the black hole surface gravity $k_E$ and $k_{TB}$ are given by:
\beqs
k_E&=&\frac{1}{A}\frac{\sigma}{m+\sigma}\frac{1}{\sqrt{(m+\sigma)(R+\sigma)}},~~~k_{TB}=\frac{1}{2A^2}\frac{1}{\sqrt{R^2-\sigma^2}}.
\eeqs
The rod directions correspond to a Taub-bolt instanton if one takes $k_{TB}\equiv\frac{1}{4n}$. Replacing now $A^2=\frac{1}{1-\alpha^2}$ and $C^2=\frac{\alpha^2}{1-\alpha^2}$ and defining $\sigma=c\kappa^2$ and $R=\kappa^2$, one is led to the same regularity condition as in the uncharged case, namely $\alpha=\pm\frac{\sqrt{1-c^2}}{2}$.

\section{Double black hole on the Taub-bolt instanton}

The aim of this section is to extend the results of the previous section and derive in closed form the general charged solution describing a static system of double black holes situated at the two turning points of the Taub-bolt geometry. To this end we shall use the four-dimensional double Reissner-Nordstr\"om solution in the parameterization given recently by Manko in \cite{Manko:2007hi}: 
\begin{equation}
\tilde{f}=\frac{\tilde{A}^{2}-\tilde{B}^{2}+\tilde{C}^{2}}{(\tilde{A}+\tilde{B})^{2}},~~~~~~e^{2\tilde{\mu}}=\frac{%
\tilde{A}^{2}-\tilde{B}^{2}+\tilde{C}^{2}}{16\sigma _{1}^{2}\sigma _{2}^{2}(\nu
+2k)^{2}r_{1}r_{2}r_{3}r_{4}},~~~~~~~\Psi=-\frac{2\tilde{C}}{\tilde{A}+\tilde{B}},  \label{Manko}
\end{equation}%
where:
\begin{eqnarray}
\tilde{A} &=&\sigma _{1}\sigma _{2}[\nu
(r_{1}+r_{2})(r_{3}+r_{4})+4k(r_{1}r_{2}+r_{3}r_{4})]-(\mu ^{2}\nu
-2k^{2})(r_{1}-r_{2})(r_{3}-r_{4}),  \notag \\
\tilde{B} &=&2\sigma _{1}\sigma _{2}[(\nu M_{1}+2kM_{2})(r_{1}+r_{2})+(\nu
M_{2}+2kM_{1})(r_{3}+r_{4})]  \notag \\
&&-2\sigma _{1}[\nu \mu (Q_{2}+\mu )+2k(RM_{2}+\mu Q_{1}-\mu
^{2})](r_{1}-r_{2})  \notag \\
&&-2\sigma _{2}[\nu \mu (Q_{1}-\mu )-2k(RM_{1}-\mu Q_{2}-\mu
^{2})](r_{3}-r_{4}),  \notag \\
\tilde{C} &=&2\sigma _{1}\sigma _{2}\{[\nu (Q_{1}-\mu )+2k(Q_{2}+\mu
)](r_{1}+r_{2})+[\nu (Q_{2}+\mu )+2k(Q_{1}-\mu )](r_{3}+r_{4})\}  \notag \\
&&-2\sigma _{1}[\mu \nu M_{2}+2k(\mu M_{1}+RQ_{2}+\mu R)](r_{1}-r_{2})
\notag \\
&&-2\sigma _{2}[\mu \nu M_{1}+2k(\mu M_{2}-RQ_{1}+\mu R)](r_{3}-r_{4}),
\end{eqnarray}%
with constants:
\begin{eqnarray}
\nu &=&R^{2}-\sigma _{1}^{2}-\sigma _{2}^{2}+2\mu
^{2},~~~~~~~k=M_{1}M_{2}-(Q_{1}-\mu )(Q_{2}+\mu ),  \notag \\
\sigma _{1}^{2} &=&M_{1}^{2}-Q_{1}^{2}+2\mu Q_{1},~~~~~~~\sigma
_{2}^{2}=M_{2}^{2}-Q_{2}^{2}-2\mu Q_{2},~~~~~~~\mu =\frac{%
M_{2}Q_{1}-M_{1}Q_{2}}{M_{1}+M_{2}+R},
\end{eqnarray}%
while $r_{i}=\sqrt{\rho ^{2}+\zeta _{i}^{2}}$, for $i=1..4$, with:
\begin{equation}
\zeta _{1}=z-\frac{R}{2}-\sigma _{2},~~~~~\zeta _{2}=z-\frac{R}{2}+\sigma
_{2},~~~~~\zeta _{3}=z+\frac{R}{2}-\sigma _{1},~~~~~\zeta _{4}=z+\frac{R}{2}%
+\sigma _{1}.
\label{ai}
\end{equation}%
This solution is parameterized by five independent parameters and describes
the superposition of two general Reissner-Nordstr\"{o}m black holes, with
masses $M_{1,2}$, charges $Q_{1,2}$ and $R$ the coordinate distance
separating them. For a detailed discussion of its properties we refer the
reader to \cite{Manko:2007hi} and the references therein. Note that, in general, the function $e^{2\tilde{\mu}}$ can be determined up to a
constant and its precise numerical value has been fixed here by allowing the
presence of conical singularities only in the portion in between the black
holes along the $\varphi $ axis. Consequently one has:
\begin{equation}
e^{2\tilde{\mu}}|_{\rho =0}=\left( \frac{\nu -2k}{\nu +2k}\right) ^{2},
\label{strutManko}
\end{equation}%
for $-R/2+\sigma _{1}<z<R/2-\sigma _{2}$ and $e^{2\tilde{\mu}}|_{\rho =0}=1$
elsewhere. 

To obtain a solution that describes two black holes sitting at the two turning points of the Taub-bolt geometry, it turns out that one has to pick the following harmonic function:
\beqs
e^{2h}&=&\sqrt{\frac{r_3+\zeta_3}{r_4+\zeta_4}}\frac{r_2+\zeta_2}{r_3+\zeta_3}\sqrt{\frac{r_1+\zeta_1}{r_2+\zeta_2}}\equiv\sqrt{\frac{(r_1+\zeta_1)(r_2+\zeta_2)}{(r_3+\zeta_3)(r_4+\zeta_4)}}.
\label{e2h}
\eeqs
The first and the third factors in the first equality correspond to the corrections associated to the black hole horizons in the four-dimensional seed solution, while the middle term corresponds to a finite rod along the $\chi$-direction, which starts at $a_3$ and ends at $a_2$. One can now easily integrate (\ref{gammap1a}) and obtain:
\beqs
e^{2\gamma}&=&\frac{2}{K_0}\left(\frac{Y_{13}Y_{14}Y_{23}Y_{24}}{r_1r_2r_3r_4Y_{12}Y_{34}}\right)^{\frac{1}{4}}.
\label{e2g}
\eeqs
Up to a constant term, the `dual' of $h$ is given by $H=\frac{1}{4}(r_3+r_4-r_1-r_2)$. Then the final five-dimensional solution is given by:
\beqs
ds^2&=&-\tilde{f}dt^2+\frac{F}{\Sigma}(d\chi+\omega d\varphi)^2+\frac{\Sigma}{G}\big[e^{2\mu}(d\rho^2+dz^2)+\rho^2 d\varphi^2\big],\nonumber\\
A_{(1)t}&=&\frac{\sqrt{3}}{2}\Psi,
\eeqs
where again the metric functions are defined as in (\ref{functions}). In order to have the right asymptotic behaviour, one has to set $A^2=C^2+1$.

\subsection{The rod structure}

Before we discuss some of its physical properties, let us consider first the rod structure of this general solution. Following the procedure given in \cite{Harmark:2004rm,Chen:2010zu}, one deduces that the rod structure is described by four turning points that divide the $z$-axis into five rods:
\begin{itemize}
\item For $z<a_4$ one has a semi-infinite spacelike with normalized direction 
\beqs
l_1=\sqrt{\frac{2\sqrt{2}}{K_0}}(0,2ACR,1).
\eeqs
\item For $a_4<z<a_3$ one has a finite timelike rod that corresponds to the first black hole horizon. Its normalized rod direction is given by $l_2=\frac{1}{k_E^{(1)}}(1,0,0)$, where 
\beqs
k_E^{(1)}=\sqrt{\frac{K_0}{2\sqrt{2}}}\frac{p_1}{A}\bigg[\left(\rho\tilde{f}^{-1}e^{\tilde{\mu}}\right)|_{\rho=0}^{(1)}\bigg]^{-\frac{3}{4}}
\eeqs
 is the surface gravity on the black hole horizon represented by this rod.
\item For $a_3<z<a_2$ one has a finite spacelike rod with normalized direction $l_3=\frac{1}{k_{TB}}(0,1,0)$, where 
\beqs
k_{TB}=\frac{1}{2A^2}\sqrt{\frac{K_0}{2\sqrt{2}}}\left(\frac{\nu+2k}{\nu-2k}\right)^{\frac{3}{4}}\frac{1}{\big[((R+\sigma_2)^2-\sigma_1^2)((R-\sigma_2)^2-\sigma_1^2)\big]^{\frac{1}{4}}}.
\eeqs
\item For $a_2<z<a_1$ one has a finite timelike rod, corresponding to the second black hole horizon. Its normalized rod direction is found to be $l_4=\frac{1}{k_E^{(2)}}(1,0,0)$, where 
\beqs
k_E^{(2)}=\sqrt{\frac{K_0}{2\sqrt{2}}}\frac{p_2}{A}\bigg[\left(\rho\tilde{f}^{-1}e^{\tilde{\mu}}\right)|_{\rho=0}^{(2)}\bigg]^{-\frac{3}{4}}
\eeqs
 is the surface gravity of the black hole horizon corresponding to this rod.
\item Finally, for $z>a_1$ one has an semi-infinite spacelike rod with normalized direction 
\beqs
l_5=\sqrt{\frac{2\sqrt{2}}{K_0}}(0,-2ACR,1).
\eeqs
\end{itemize}

Here we defined the following quantities:
\beqs
p_1&=&\left(\frac{\sigma_1}{(R+\sigma_1)^2-\sigma_2^2}\right)^{\frac{1}{4}},~~~~~p_2=\left(\frac{\sigma_2}{(R+\sigma_2)^2-\sigma_1^2}\right)^{\frac{1}{4}},
\label{pi}
\eeqs
while for each black hole horizon one has \cite{Manko:2008gb}:
\beqs
\left(\rho\tilde{f}^{-1}e^{\tilde{\mu}}\right)|_{\rho=0}^{(1)}&=&\frac{\big[(R+M_1+M_2)(M_1+\sigma_1)-Q_1(Q_1+Q_2)\big]^2}{\sigma_1[(R+\sigma_1)^2-\sigma_2^2]},\nonumber\\
\left(\rho\tilde{f}^{-1}e^{\tilde{\mu}}\right)|_{\rho=0}^{(2)}&=&\frac{\big[(R+M_1+M_2)(M_2+\sigma_2)-Q_2(Q_1+Q_2)\big]^2}{\sigma_2[(R+\sigma_2)^2-\sigma_1^2]}.
\label{rhofmu}
\eeqs
Let us pick now $K_0=2\sqrt{2}$ and further define the nut charge $n=ACR$. Then one obtains the rod structure of the Taub-bolt instanton if one sets:
\beqs
C&=&A\left(\frac{\nu-2k}{\nu+2k}\right)^{\frac{3}{4}}\frac{\big[((R+\sigma_2)^2-\sigma_1^2)((R-\sigma_2)^2-\sigma_1^2)\big]^{\frac{1}{4}}}{2R},
\label{cond}
\eeqs
such that the rod directions outside the black hole horizons become $l_1=(0,2n,1)$, $l_3=(0,4n,0)$ and $l_5=(0,-2n,1)$, characteristic of the Taub-bolt instanton background. Finally, to ensure regularity of the background geometry, the following identifications of the coordinates $(\chi,\varphi)$ have to be made \cite{Chen:2010ih}:
\beqs
(\chi,\varphi)\rightarrow (\chi+4n\pi,\varphi+2\pi),~~~(\chi,\varphi)\rightarrow(\chi+8n\pi,\varphi).
\eeqs
Together with the asymptotic condition $A^2-C^2=1$, the relation (\ref{cond}) completely fixes the values of the constants $A$ and $C$ in terms of the physical parameters appearing in the final solution.

\subsection{Particular cases}

Consider first the limit in which the second black hole is absent. For future convenience, perform a shift $z\ra z-R/2$ of the $z$ coordinate such that one centers on the horizon of the first black hole. This amounts to taking $M_2=Q_2=0$, that is $r_1=r_2$. Since in this case the four-dimensional seed reduces to the single Reissner-Nordstr\"om black hole, it should be clear that using the expressions of $e^{2h}$ and $e^{2\gamma}$ in (\ref{e2h}) and (\ref{e2g}) one recovers the charged single black hole solution from the previous section.

Another limit of interest is the extremal limit. This can be achieved by setting $Q_i=\epsilon M_i$, where $\epsilon=\pm 1$, for each $i=1,2$. This leads to $\sigma_1=\sigma_2=k=\mu=0$ and, in consequence, in this limit $r_1=r_2$ while $r_3=r_4$. Now, once we have $\sigma_1=\sigma_2=0$ then (\ref{Manko}) becomes:
\begin{eqnarray}
\tilde{f}_e&=&\left(1+\frac{M_1}{r_3}+\frac{M_2}{r_1}\right)^{-2},~~~~~~~
e^{2\tilde{\mu}}|_{e}=1,~~~~~A_{(1)t}=-\frac{\sqrt{3}}{2}\left(1+\frac{M_1}{r_3}+\frac{M_2}{r_1}\right)^{-1}.
\label{extremal4dManko}
\end{eqnarray}
Finally, if $r_1=r_2$ and $r_3=r_4$ then:
\beqs
e^{2h}&=&\frac{r_1+\zeta_1}{r_3+\zeta_3},~~~~e^{2\gamma}=\frac{Y_{13}}{2r_1r_3},~~~\omega=2AC(r_3-r_1),~~~\Sigma=A^2-C^2\left(\frac{r_1+\zeta_1}{r_3+\zeta_3}\right)^2,
\label{back}
\eeqs 
where $r_1=\sqrt{\rho^2+(z-\frac{R}{2})^2}$ and $r_3=\sqrt{\rho^2+(z+\frac{R}{2})^2}$. Gathering up all these results, the solution describing a pair of extremal black holes sitting at the two turning points of the Taub-bolt instanton can be written as:
\beqs
ds^2&=&-\frac{dt^2}{\left(1+\frac{M_1}{r_3}+\frac{M_2}{r_1}\right)^{-2}}+\left(1+\frac{M_1}{r_3}+\frac{M_2}{r_1}\right)\bigg[\frac{e^{2h}}{\Sigma}(d\chi+\omega d\varphi)^2+\frac{\Sigma}{ e^{2h}}\big[e^{2\gamma}(d\rho^2+dz^2)+\rho^2d\varphi^2\big]\bigg],\nonumber\\
A_{(1)t}&=&-\frac{\sqrt{3}}{2}\left(1+\frac{M_1}{r_3}+\frac{M_2}{r_1}\right)^{-1}.
\label{ext}
\eeqs
To show that the background geometry corresponds indeed to the regular Taub-bolt instanton, we shall perform the following coordinate transformations \cite{Chen:2010zu}:
\beqs
\rho&=&\sqrt{(r-m)^2-\frac{R^2}{4}}\sin\theta,~~~~~z=(r-m)\cos\theta.
\eeqs
with $\frac{R}{2}\equiv\sqrt{m^2-n^2}$, where $m$ and $n$ are the mass and nut charge of the four-dimensional background geometry. After imposing the asymptotic condition $A^2=C^2+1$, it turns out that the background geometry  described by (\ref{back}) can be written in the Taub-NUT-like form:
\beqs
ds^2_{(4)}&=&f(r)(d\chi+2n\cos\theta d\varphi)^2+\frac{dr^2}{f(r)}+(r^2-n^2)(d\theta^2+\sin^2\theta d\varphi^2),\nonumber\\
f(r)&=&\frac{r^2-2mr+n^2}{r^2-n^2},
\eeqs
if one picks $C^2=\frac{m-\sqrt{m^2-n^2}}{2\sqrt{m^2-n^2}}$. Recall now that according to (\ref{cond}) one has to take $C=\frac{A}{2}$ to have the right rod structure of the final five dimensional solution. Solving the constraints one obtains $A^2=\frac{4}{3}$ and $C^2=\frac{1}{3}$, while the nut charge becomes $n=\frac{A^2R}{2}=\frac{2R}{3}$. Keeping in mind the above expression of $R$ in terms of $m$ and $n$, it is easy to see that this condition amounts to taking $m=\frac{5|n|}{4}$, which is in fact the regularity condition of the Taub-bolt instanton. In conclusion, we have shown that the extremal solution (\ref{ext}) correctly describes a pair of extremal black holes in equilibrium, situated at the two turning points of the Taub-bolt instanton.

\subsection{Conserved charges and thermodynamics}

The asymptotic region is found after performing the coordinate transformations:
\beqs
\rho=r\sin\theta,~~~~~~z=r\cos\theta.
\eeqs
and taking the limit $r\rightarrow\infty$. In general, the conserved charges are encoded in the asymptotic expansions of the metric functions. For geometries that asymptote to that of the KK monopole, to compute the conserved charges one possibility is to use the counter-terms method, as described for instance in \cite{Mann:2005cx} or \cite{Kleihaus:2009ff}.

One finds the total conserved mass and the gravitational tension to be given by:
\beqs
{\cal M}&=&\frac{L}{4G}\big[3(M_1+M_2)+R(1+2C^2)\big],~~~~~{\cal T}=\frac{R(1+2C^2)}{2G},
\eeqs
where $G$ is the gravitational constant in five dimensions, while $L=8\pi n$ is the length of the $\chi$ circle at infinity. The total charge is computed by using Gauss' formula, with the result:
\beqs
{\cal Q}&=&\frac{\sqrt{3}(Q_1+Q2)L}{4G}.
\eeqs
One can also compute the individual charges, for each black hole horizon and one finds:
\beqs
{\cal Q}_i&=&\frac{\sqrt{3}Q_iL}{4G}.
\eeqs
Note that the total charge is the sum of the individual black hole charges, as expected: ${\cal Q}={\cal Q}_1+{\cal Q}_2$. The electric potential on each black hole horizon can also be computed 
\beqs
\Phi_H^i&=&-A_{(1)t}|_{horizon}=\sqrt{3}\left(\frac{M_i-\sigma_i}{Q_i}\right).
\eeqs

One can also relate the five dimensional areas of each black hole horizon to the corresponding areas of the black hole horizons in the four dimensional seed solution. More precisely, in four dimensions, the area of each black hole horizon can be expressed as  \cite{Manko:2008gb}:
\beqs
A_{(4)}^i&=&4\pi\sigma_i\left(\rho\tilde{f}^{-1}e^{\tilde{\mu}}\right)|_{\rho=0}^i,
\eeqs
where for each black hole horizon the quantities $\left(\rho\tilde{f}^{-1}e^{\tilde{\mu}}\right)|_{\rho=0}^i$ are given in (\ref{rhofmu}). 
For the final five-dimensional solution the area of each black hole horizon can be written as:
\beqs
A_{(5)}^i&=&4\pi \sigma_iL\big[(\rho\tilde{f}^{-1}e^{\tilde{\mu}})|_{\rho=0}^i\big]^{\frac{3}{4}}\left((\rho^{\frac{1}{2}}e^{2\gamma-2h}\Sigma)|_{\rho=0}^i\right)^{\frac{1}{2}}.
\label{area}
\eeqs
Here $\Sigma|_{\rho=0}=\Sigma_0=1+C^2$, while near each black hole horizon one can expand:
\beqs
e^{2h-2\gamma}|_i&=&(p_i)^2\sqrt{\rho}+{\cal O}(\rho),
\eeqs
where the constants $p_i$ have been previously defined in (\ref{pi}). Using (\ref{area}) one finds the particularly simple expressions:
\beqs
A_{(5)}^1&=&4\pi \sqrt{1+C^2}L\left(\frac{\big[(R+M_1+M_2)(M_1+\sigma_1)-Q_1(Q_1+Q_2)\big]^{3}}{(R+\sigma_1)^2-\sigma_2^2}\right)^{\frac{1}{2}},\nonumber\\
A_{(5)}^2&=&4\pi \sqrt{1+C^2}L\left(\frac{\big[(R+M_1+M_2)(M_2+\sigma_2)-Q_2(Q_1+Q_2)\big]^{3}}{(R+\sigma_2)^2-\sigma_1^2}\right)^{\frac{1}{2}}.
\eeqs
Finally, the Hawking temperature for each black hole horizon is computed using its definition in terms of the surface gravity $k_i$. Specifically, one finds:
\beqs
k_i&=&\frac{p_i}{\sqrt{\Sigma_0}}\bigg[\left(\rho\tilde{f}^{-1}e^{\tilde{\mu}}\right)|_{\rho=0}^i\bigg]^{-\frac{3}{4}},
\eeqs
such that the Hawking temperature of each black hole is given by $T_i=\frac{k_i}{2\pi}$. As a check of our computation, note that the surface gravities computed here agree with those appearing in the normalization of the rod directions, as expected. Note now the simple relation:
\beqs
\frac{A_{(5)}^ik_i}{8\pi G}&=&\frac{L\sigma_i}{2G},
\eeqs
which can be used to prove a simple Smarr relation for this system. To this end, one has to compute the so-called Komar mass:
\beqs
M_{K}&=&-\frac{1}{16\pi G }\frac{3}{2}\int_{S}\alpha,
\label{MK}
\eeqs
where $S$ is the boundary of any spacelike hypersurface and:
\begin{equation}
\alpha _{\mu \nu \rho }=\epsilon _{\mu \nu \rho \sigma \tau }\nabla ^{\sigma
}\xi ^{\tau }\ ,
\end{equation}
with the Killing vector $\xi =\partial /\partial t$. This quantity is a measure of the mass contained in $S$, and if we take $S$ to be the squashed three-sphere at infinity enclosing both horizons then (\ref{MK}) gives the total Komar mass of the system:
\beqs
M_K&=&\frac{3(M_1+M_2){\cal L}}{4G},
\eeqs
 while the Komar mass of each individual black hole is obtained by performing the above integration at the respective black hole horizon $M_{Komar}^i=\frac{3L\sigma_i}{4G}$. At this point, let us note that the relation $2M_K=2{\cal M}-{\cal T}{\cal L}$ is satisfied for this multi-black hole configuration. Putting all things together, one arrives to the following Smarr relation for the double black hole system on the Taub-bolt instanton:
\beqs
2{\cal M}-{\cal T}L=2M_K=3\left(\frac{A_{(5)}^1k_1}{8\pi G}+\frac{A_{(5)}^2k_2}{8\pi G}\right)+2\Phi^1{\cal Q}_1+2\Phi^2{\cal Q}_2.
\eeqs
Also, one can readily check the individual Smarr relayions for each black hole component:
\beqs
2M_{K}^{(i)}=3\left(\frac{A_{(5)}^ik_{i}}{8\pi G}\right)+2\Phi^i{\cal Q}_i,
\eeqs
where $M_{K}^{(i)}=\frac{3M_i{\cal L}}{4G}$. Thus one can regard $M_{K}^{(i)}$ as the individual mass of each black object, containing an electromagnetic contribution apart from the Komar piece.

\section{Conclusions}

In this work we made use of a solution generating technique to derive a new exact solution describing a general charged double black hole configuration on the Taub-bolt instanton. This method has been previously used in \cite{Stelea:2009ur,Stelea:2011fj} to obtain exact solutions describing general configurations of charged black holes in five dimensional spaces whose asymptotic geometry resembles that of the KK magnetic monopole. In the first part of this paper, in section $3$ we were able to re-derive a previously known solution describing a static charged black hole sitting on one of the two turning points of the Taub-bolt gravitational instanton. In section 4 we generalized this construction by using the four dimensional double Reissner-Nordstr\"om solution as seed \cite{Manko:2007hi}. We showed that one can pick the parameters such that the conical singularities are avoided and the rod structure of the final solution resembles that of the Taub-bolt instanton. Finally, by using a counter-terms method we computed the conserved charges and showed that a Smarr relation is satisfied in this case.

As a general feature, in four and five dimensions, solutions describing multi-horizon objects are plagued by unavoidable conical singularities. From a physical point of view, the presence of these conical singularities is to be expected since they are needed to balance the gravitational attraction forces in between the black holes and their electromagnetic interaction, in order to keep the system static. Previous work \cite{Chng:2008sr,Stelea:2011jm,Stelea:2009ur,Stelea:2011fj} showed that in absence of rotation the electromagnetic interaction is not strong enough to balance the gravitational attraction in between the black holes and the conical singularities cannot be avoided. On the other hand, even if the solutions containing conical singularities appear to be singular at those points, such systems can still have a well-defined gravitational action \cite{Gibbons:1979nf,Costa:2000kf,Herdeiro:2010aq,Herdeiro:2009vd}. This means that such multi-black hole solutions with conical singularities might still admit a reasonable and well-defined thermodynamic description. For spaces with KK asymptotics, this has been explicitly checked in \cite{Stelea:2011fj}. In the present context, it turns out that the double black hole on the Taub-bolt instanton can be equilibrated, even if the black holes are non-extremal. To our knowledge, this is the first example of a general charged static double-black hole system in equilibrium in five dimensions, without using Kaluza-Klein bubbles to balance the system. 

As avenues for further work, it would be interesting to consider more general configurations, containing black holes and black rings. For example, one should be able to construct a solution describing a black ring on the Taub-bolt instanton and study its properties. Work on this subject is currently in progress and it will be reported elsewhere \cite{BRonBOLT}.

\vspace{10pt}

{\Large Acknowledgements}

The work of C. S. was financially supported by POSDRU through the POSDRU/89/1.5/S/49944 contract.

\end{document}